\documentclass[prd,twocolumn]{revtex4}
\usepackage{dcolumn}
\usepackage{array}
\usepackage{multirow}
\usepackage{booktabs}
\usepackage{graphicx}
\usepackage{amssymb}
\usepackage{bm}
\usepackage{setspace}
\usepackage{hyperref}
\usepackage{epstopdf}
\usepackage{color}
\usepackage{mathrsfs}
\usepackage{amsmath,amssymb,amsthm}
\begin{document}
\title{The multi-feature universe: large parameter space cosmology and the swampland}
\author{Deng Wang}
\email{cstar@mail.nankai.edu.cn}
\affiliation{Department of Astronomy, School of Physics and Astronomy, Shanghai Jiao Tong University, Shanghai 200240, China}

\begin{abstract}
Under the belief that the universe should be multi-feature and informative, we employ a model-by-model comparison method to explore the possibly largest upper bound on the swampland constant $c$. Considering the interacting quintessence dark energy as the comparison model, we constrain the large parameter space interacting dark energy model, a 12-parameter extension to the $\Lambda$CDM cosmology, in light of current observations. We obtain the largest $2\sigma$ ($3\sigma$) bound so far, $c\lesssim1.62$ $(1.94)$, which would allow the existences of a number of string theory models of dark energy such as 11-dimensional supergravity with double-exponential potential, $O(16)\times O(16)$ heterotic string and some Type II string compactifications. For inflationary models with concave potential, we find the $2\sigma$ ($3\sigma$) bound $c\lesssim0.13$ $(0.14)$, which is still in strong tension with the string-based expectation $c \sim \mathcal{O}(1)$. However, combining Planck primordial non-Gaussianity with inflation constraints, it is interesting that the Dirac-Born-Infeld inflation with concave potential gives the $2\sigma$ ($3\sigma$) bound $c\lesssim0.53$ $(0.58)$, which is now in a modest tension with the swampland conjecture.

\end{abstract}
\maketitle

\section{Introduction}
For a long time, in modern cosmology, there are two main hot topics, i.e., early-time inflation and late-time dark energy (DE). For the very early universe, inflation, which is a hypothetical period of quasi-de Sitter exponential expansion, can give a solution to several important problems in the standard Big Bang paradigm such as the lack of relic monopoles, flatness, homogeneity and the horizon problem \cite{1,2,3,4,5,6,7,8}. Moreover, inflation can also explain the primordial density perturbations derived from the observation of cosmic microwave background (CMB) anisotropies \cite{9,10,11,12}. For the late universe, since the cosmic acceleration was discovered by two Type Ia supernovae (SNe Ia) groups \cite{13,14}, the existence of DE has also been verified during the past two decades by many independent cosmological probes at various cosmological scales such as the CMB radiation \cite{15,16}, baryon acoustic oscillations (BAO) \cite{17,18}, X-ray clusters \cite{19,20} and weak gravitational lensing \cite{21}. Meanwhile, cosmologists have established the standard DE scenario, i.e., the $\Lambda$-cold dark matter ($\Lambda$CDM) model. Most recently, Planck-2018 final release \cite{22} with improved measurement of the reionization optical depth has confirmed, once again, the validity of the simple 6-parameter $\Lambda$CDM cosmology in describing the evolution of the universe, although there still exist the Hubble constant ($H_0$) tension and matter fluctuation amplitude ($\sigma_8$) tension. Therefore, current standard cosmological paradigm embedded in the framework of general relativity (GR) should be `` inflation$+$$\Lambda$CDM ''.

However, GR cannot be the ultimate theory to characterize such a realistic and tremendous universe from cosmological scales to extremely small scales. A common viewpoint is that GR could be the low energy limit of a well-motivated high energy UV-complete theory, where the behavior of inflaton field and DE phenomenon can be usually captured by the effective field theory (EFT) originated from its low energy limit. String theory, a unified theory combining the standard model of particle physics with gravity, naturally emerges as a candidate for such a UV-complete theory and has received much interest. String theory provides a vast landscape of vacua and such a landscape is believed to lead to consistent EFTs, which are surrounded by the so-called swampland, a region wherein inconsistent semi-classical EFTs inhabit. Consequently, finding a theoretical boundary or constructing a set of conditions to distinguish the consistent EFTs from inconsistent ones lying in the swampland is a very urgent task. Weak gravity conjecture \cite{23} and recently proposed swampland criteria \cite{24,25} have attempted to address this important issue and received substantial interest. It is noteworthy that, as is well known, Minkowski and Anti-de Sitter solutions in string landscape is easy to be obtained, but the de Sitter ones are extremely difficult to be found \cite{26,27}. Hence, it is reasonable to guess that de Sitter vacua may reside in the swampland not in the landscape. In light of this, one of two swampland criteria has made the de Sitter vacua to be part of the swampland \cite{28}.

In this study, we mainly focus on the cosmological implications of the swampland conjecture (\emph{SC}) \cite{29}, which is expressed as:

$\star$ \emph{SC}1: The scalar field net excursion in reduced Planck units should satisfy the bound
\vspace{0.0001cm}
\begin{equation}
\frac{|\Delta\phi|}{M_p}<\Delta\sim \mathcal{O}(1), \label{1}
\end{equation}

$\star$ \emph{SC}2: The gradient of the scalar field potential is bounded by
\vspace{0.0001cm}
\begin{equation}
M_p\frac{|V'|}{V}>c \sim \mathcal{O}(1), \label{2}
\end{equation}

where both $\Delta$ and $c$ are positive constants of order unity, the prime denotes the derivative with respect to the scalar field $\phi$, and $M_p=1/\sqrt{8\pi G}$ is the reduced Planck mass.

Recently, when applying these two \emph{SCs} to observational cosmology, Agrawal {\it et al}. \cite{29} find that: (i) \emph{SCs} are in very strong tension with observationally living plateau inflationary models by requiring $c\lesssim$0.02 and $\Delta\gtrsim5$; (ii) Although \emph{SC}2 disfavors clearly the $\Lambda$CDM model, quintessence DE model can be made compatible with two \emph{SCs} by demanding $c<0.6$ and $c<3.5\Delta$. Subsequently, for DE, Heisenberg {\it et al}. \cite{30} give current $3\sigma$ constraint with $c\lesssim1.35$ via a Finsher matrix analysis, and for inflation, kinney {\it et al}. \cite{31} derive out the relation between the second slow-roll parameter and $c$ and find a bound $c\lesssim0.1$ based on their constraint. Interestingly, Akrami {\it et al}. \cite{32} obtain the $3\sigma$ bound $c\lesssim1.01$ by constraining a 5-parameter quintessence DE model with current data, and they rule out recently proposed alternative string theory models of DE at least at the $3\sigma$ confidence level. Note that their $2\sigma$ bound $c\lesssim0.78$ \cite{32} is a little larger than that $c<0.6$ obtained by Agrawal {\it et al} \cite{29}.

In light of the above results, we aim at giving the possibly largest upper bound on $c$ based on currently available cosmological data.

\section{Analysis}
In \cite{29}, there are three premises to obtain the bound $c<0.6$ from DE analysis using a model-by-model comparison method: (i) Choosing the standard 8-parameter Chevallier-Polarski-Linder (CPL) cosmology as the fiducial model to fit data; (ii) Choosing the quintessence DE cosmology as the comparison model; (iii) Choosing the simplest exponential potential for quintessence field.

We argue that these three assumptions should be changed at least in light of the following two viewpoints. Firstly, such the realistic and tremendous universe must be extremely informative and multi-feature, and we cannot use a oversimplified 6-parameter ($\Lambda$CDM) or 8-parameter model to characterize it. For example, in practice, we have no reason to fix the sum of three active neutrinos to be $\Sigma_\nu=0.06$ eV and assume the equation of state of DE to be $\omega=-1$. Secondly, actual cosmological tension must be stable when decreasing or increasing the number of dimension of parameter space. The tension generated by ignoring the physically related degrees of freedom should not be the real one.

Under the belief that the universe must be multi-feature and informative, for the late universe, we attempt to explore the possibly largest upper limit of $c$ in a physically large parameter space cosmology. Specifically, we choose a 18-parameter space scenario as the fiducial model by considering a interaction between dark matter (DM) and DE. Meanwhile, we choose the coupled quintessence cosmology as the comparison model, where quintessence DE interacts with DM in the dark sector. As for the scalar potential, we still choose the exponential one. The effects of modifications of potential on the limit of $c$ have been studied in \cite{32,33}.

Based on current B-mode constraint, in \cite{29}, obtaining the bound $c\lesssim0.09$ from inflation analysis has a main premise, namely using a 7-parameter model ($\Lambda$CDM plus the tensor-to-scalar ratio) as the fiducial one. Similarly, for the very early universe, we also use the 18-parameter space cosmology as the fiducial model to implement the analysis about the upper bound on $c$.

\begin{figure}
	\centering
	\includegraphics[scale=0.4]{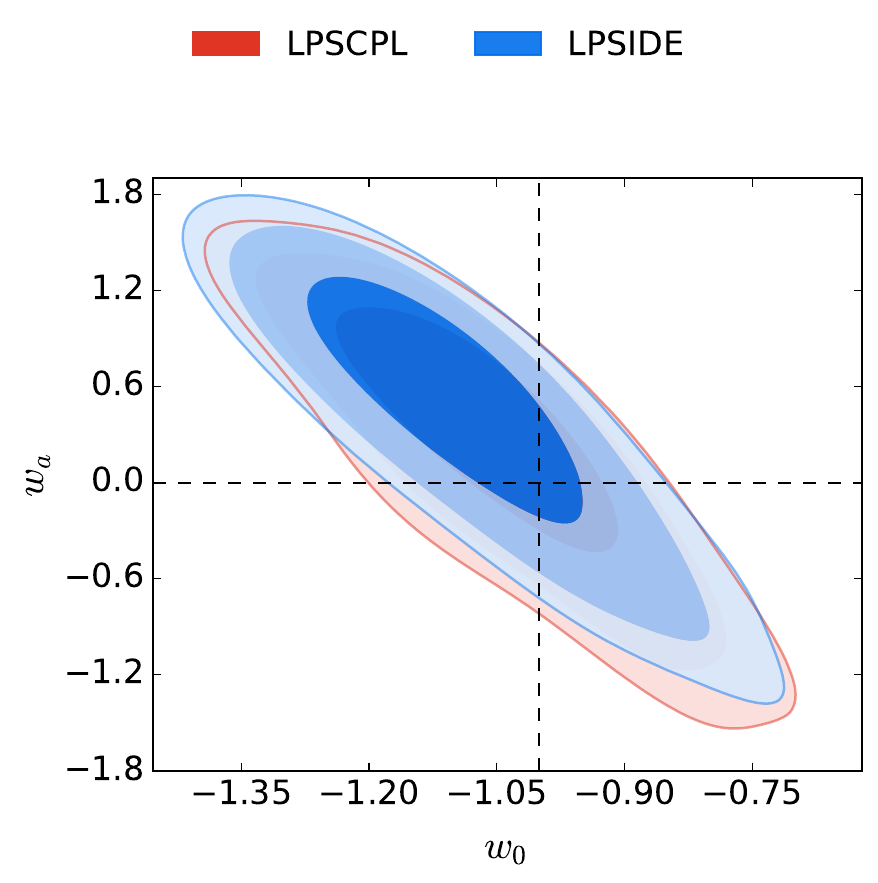}
	\caption{Marginalized $1\sigma$ ($68\%$), $2\sigma$ ($95\%$) and $3\sigma$ ($99\%$) constraints on the CPL parameter pair ($\omega_0$, $\omega_a$) using the combined datasets TLLBP are shown in the LPSCPL and LPSIDE models, respectively. The dashed lines indicate the point corresponding to the $\Lambda$CDM model. }\label{f1}
\end{figure}

\begin{table}[h!]
	\renewcommand\arraystretch{1.5}
	\caption{The $2\sigma$ ($3\sigma$) upper bound on the constant $c$ when $Q=0.001$ in the IQDE comparison model are shown in two fiducial models using the combined constraint TLLBP.}
	\begin{tabular}{l c }
		\hline
		\hline
		Model        & $2\sigma$ ($3\sigma$) bound on $c$                      \\
		\hline
		LPSCPL       &1.60 (1.93)                                        \\
		LPSIDE       &1.62 (1.95)                                       \\

		\hline
		\hline
	\end{tabular}
	\label{t1}
\end{table}

\begin{figure}
	\centering
	\includegraphics[width=7cm,height=6cm]{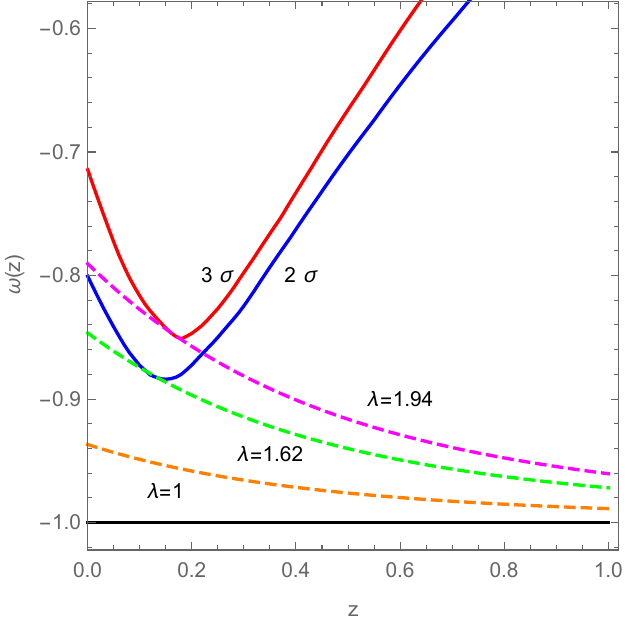}
	\caption{The red and blue solid lines are current $1\sigma$ and $2\sigma$ upper bounds on the reconstructed EoS of DE using the combined constraint TLLBP in the LPSIDE model, respectively. The magenta, green and orange dashed lines are the reconstructed EoS of DE from the IQDE model when $\lambda=1.94,\,1.62,\,1$, respectively. The black solid line is the $\Lambda$CDM model. }\label{f2}
\end{figure}

\begin{figure}
	\centering
	\includegraphics[width=8cm,height=7cm]{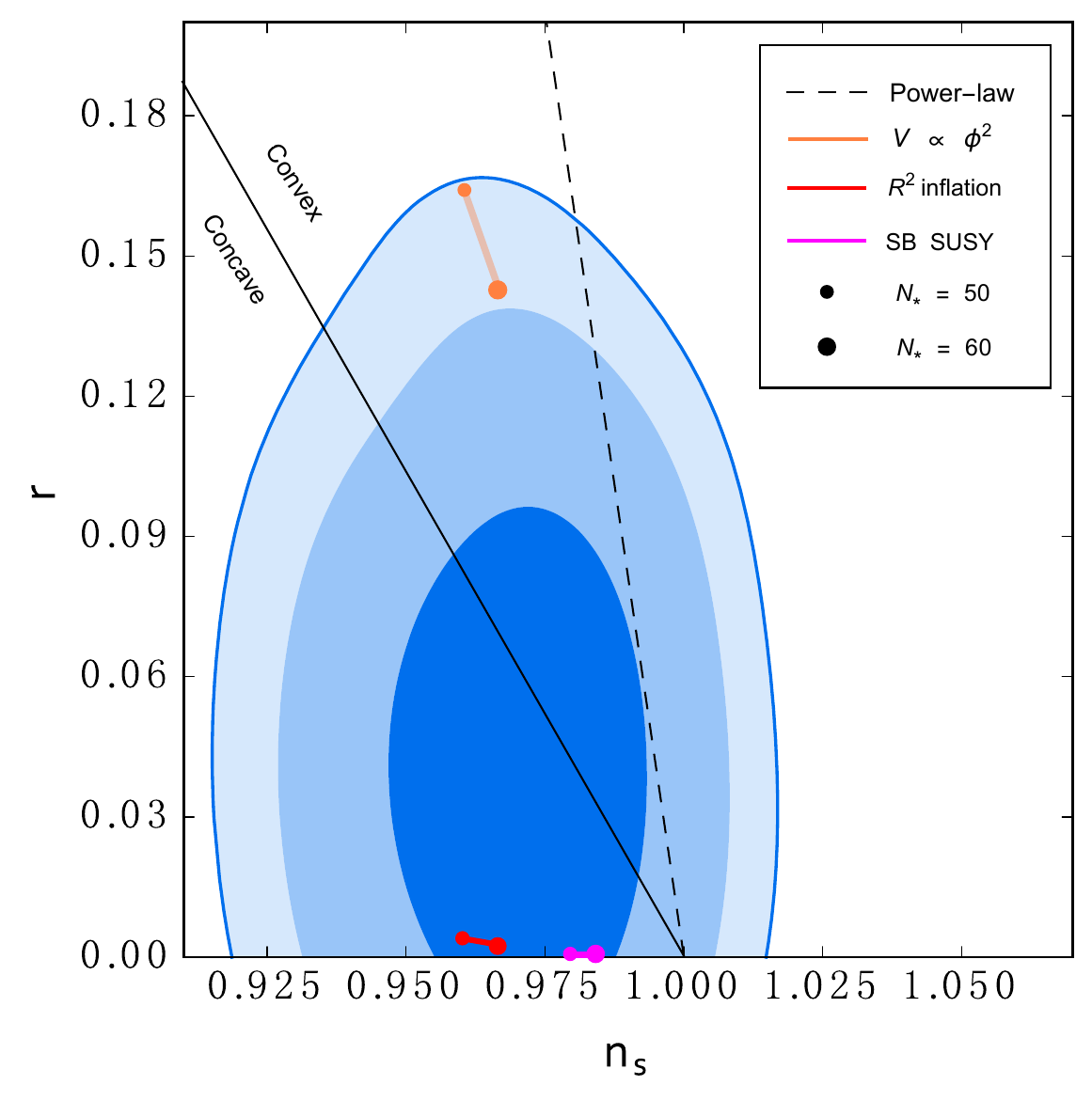}
	\caption{Marginalized $1\sigma$ ($68\%$), $2\sigma$ ($95\%$) and $3\sigma$ ($99\%$) constraints on the parameters $n_s$ and $r$ calculated at $k=0.002$ Mpc$^{-1}$ using the combined datasets TLLBK, compared to the theoretical predictions of selected inflationary models, are shown in the LPSIDE model. }\label{f3}
\end{figure}
\begin{figure*}
	\centering
	\includegraphics[scale=0.2]{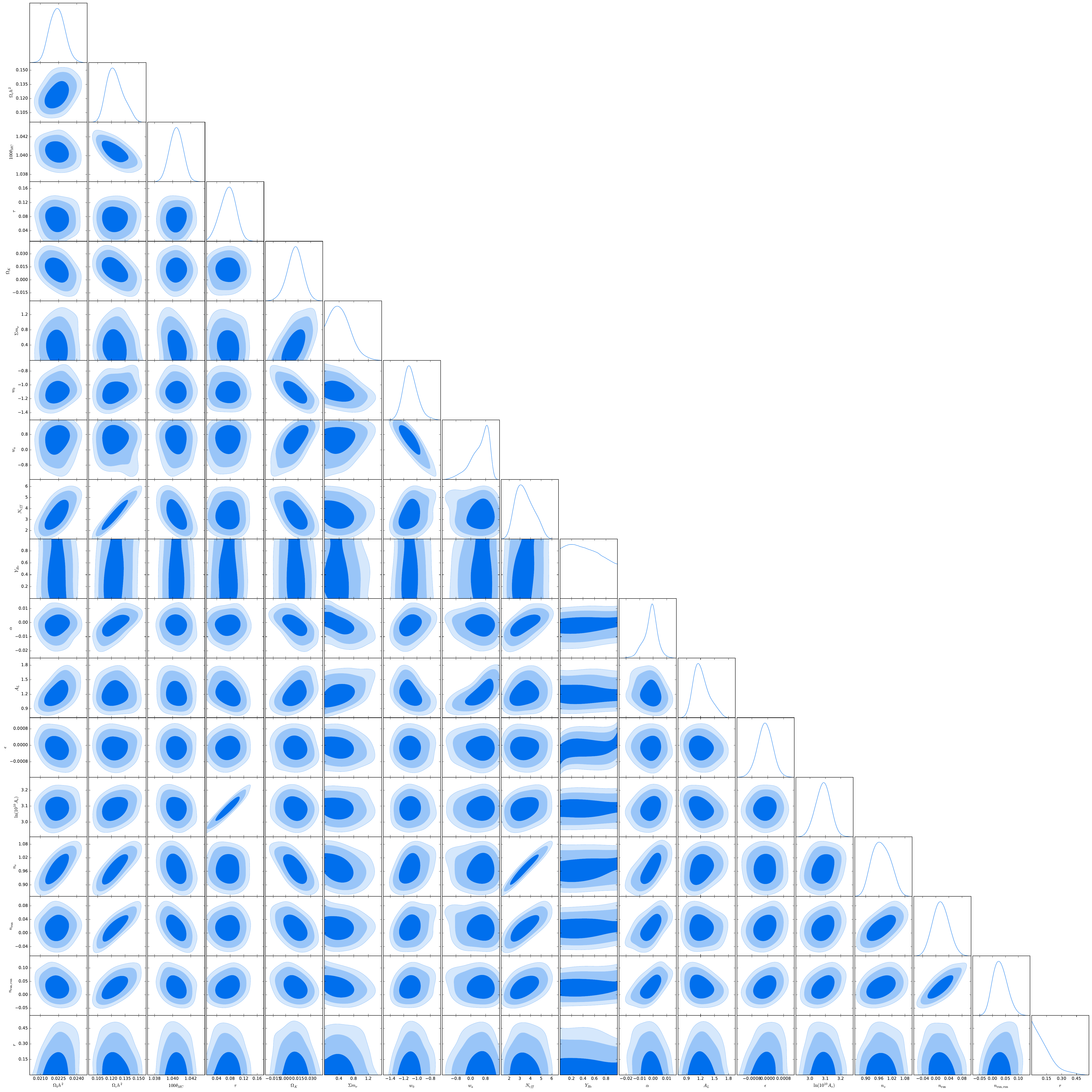}
	\caption{Marginalized $1\sigma$, $2\sigma$ and $3\sigma$ constraints on on the model parameters of the LPSIDE model using the combined datasets TLLBP. }\label{f4}
\end{figure*}
\begin{figure}
	\centering	\includegraphics[scale=0.6]{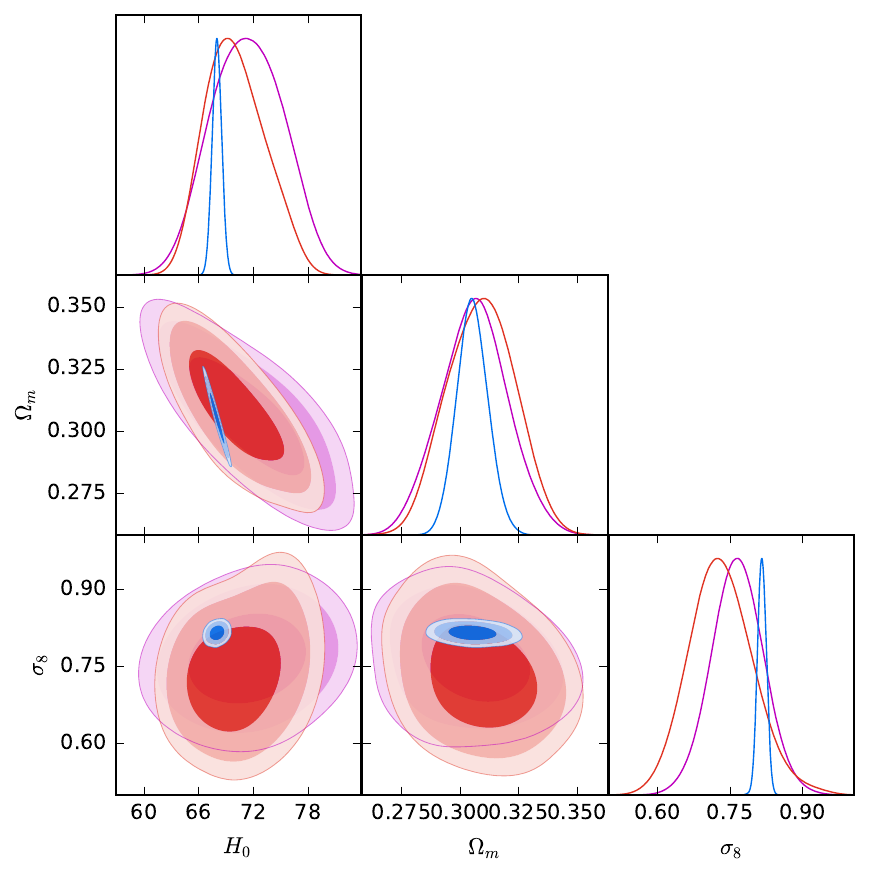}
	\caption{Marginalized $1\sigma$, $2\sigma$ and $3\sigma$ constraints on the derived parameters Hubble constant $H_0$, matter density $\Omega_m$ and matter fluctuation amplitude $\sigma_8$ in the $\Lambda$CDM (blue), LPSCPL (pink) and LPSIDE (red) models using the combined datasets TLLBP. }\label{f5}
\end{figure}

\section{Models and Methods}
As the comparison model, we consider the interacting quintessence dark energy (IQDE) scenario, whose Friedmann equation is written as $3H^2=\rho_\phi+\rho_{bm}+\rho_{dm}+\rho_r$, where $H$ is the Hubble parameter, and $\rho_\phi$, $\rho_{bm}$, $\rho_{dm}$ and $\rho_r$ are energy densities of quintessence field, baryonic matter, DM and radiation, respectively. Furthermore, we have $\dot{\rho_\phi}+3H(\rho_\phi+P_\phi)=-\dot{\rho_{dm}}-3H\rho_{dm}=-Q\rho_{dm}\dot{\phi}$, where $Q$ is the interaction between DM and DE, $P_\phi$ is the pressure of field $\phi$ and the dot denotes the derivative with respect to the cosmic time. The trajectory of field $\phi$ can be parameterized by a set of dynamical variables $x_1=\dot{\phi}/\sqrt{6}H$, $x_2=\sqrt{V}/\sqrt{3}H$, $x_3=\sqrt{\rho_{bm}}/\sqrt{3}H$ and $x_4=\sqrt{\rho_r}/\sqrt{3}H$. We use $M_p^2=1$ throughout this study. Subsequently, considering the case of exponential potential $V(\phi)=V_0e^{\lambda\phi}$, the equations of motion of this system can be
\vspace{0.0001cm}
conveniently shown as
\begin{eqnarray}
\frac{dx_1}{dN}&=&-3x_1-\frac{\sqrt{6}}{2}\lambda x_2^2-x_1\frac{1}{H}\frac{dH}{dN} \label{3} \\
               & &-\frac{\sqrt{6}}{2}Q(1-x_1^2-x_2^2-x_3^2-x_4^4),         \nonumber \\
\frac{dx_2}{dN}&=&\frac{\sqrt{6}}{2}\lambda x_1x_2-x_2\frac{1}{H}\frac{dH}{dN},  \label{4}     \\
\frac{dx_3}{dN}&=&-\frac{3}{2}x_3-x_3\frac{1}{H}\frac{dH}{dN},                   \label{5}      \\
\frac{dx_4}{dN}&=&-2x_4-x_4\frac{1}{H}\frac{dH}{dN},                           \label{6}
\end{eqnarray}
where the number of e-foldings $N=\mathrm{ln}\,a$ and $(1/H)(\mathrm{d}H/\mathrm{d}N)=-(3x_1^2-3x_2^2-3x_3^2+x_4^2+3)/2$. Note that the ratio between the slope of potential in \emph{SC}2 and the potential can now be expressed as $|V'|/V=\lambda>c \sim \mathcal{O}(1)$, which indicates that we can use the largest value of $\lambda$ to determine the bound on the swampland constant $c$. 

As the fiducial model for DE and inflation analysis, our 18-parameter scenario consists of extended 17-parameter CPL plus one interaction parameter $\epsilon$, which represents modified expansion rate of DM characterized by $\rho_{dm}=\rho_{dm0}(1+z)^{3-\epsilon}$ \cite{34}, where $\rho_{dm0}$ is present energy density of DM and $z$ is redshift. The corresponding parameter space is $\{ \Omega_bh^2, \Omega_ch^2, 100\theta_{MC}, \tau, \mathrm{ln}(10^{10}A_s), n_s, \omega_0, \omega_a, \epsilon, \Omega_K, \Sigma m_\nu, \\ N_{eff}, r, Y_{He}, A_L, n_{run}, n_{run,run}, \alpha \}$, where $\Omega_bh^2$ and $\Omega_ch^2$ are present baryon and DM densities, $\theta_{MC}$ is angular scale of acoustic horizon $\theta$ at decoupling, $\tau$ is reionization optical depth, $\mathrm{ln}(10^{10}A_s)$ and $n_s$ are amplitude and spectral index of scalar power spectrum, $\omega_0$ and $\omega_a$ are two parameters of CPL parametrization, $\epsilon$ is modified expansion rate of DM, $\Omega_K$ is cosmic curvature, $\Sigma m_\nu$ is sum of three active neutrinos, $N_{eff}$ is effective number of relative species, $r$ is tensor-to-scalar ratio calculated at pivot scale $k=0.002$ Mpc$^{-1}$, $Y_{He}$ is primordial helium abundance, $A_L$ is consistency parameter of lensing spectrum, $n_{run}$ is running of $n_s$, $n_{run,run}$ is running of running of $n_s$, and $\alpha$ is correlated matter isocurvature amplitude, respectively. $h$ is related to $H_0$ by $H_0/h\equiv 100$ km s$^{-1}$ Mpc$^{-1}$. Hereafter we refer to our 18-parameter scenario as large parameter space interacting dark energy (LPSIDE). Correspondingly, 17-parameter CPL is denoted as LPSCPL.

To perform the standard Bayesian analysis, we have modified the publicly Markov Chain Monte Carlo code $\mathbf{CosmoMC}$ \cite{35} and Boltzmann code $\mathbf{CAMB}$ \cite{36} for LPSIDE. We choose flat priors for all the parameters and marginalize the foreground nuisance parameters provided by Planck. We use CMB data including Planck-2015 temperature, polarization and lensing (TTTEEE+lowP+lensing) \cite{22}, BAO data including 6dFGS \cite{37}, SDSS-MGS \cite{38} and consensus measurement from BOSS DR12 combined sample \cite{39}, the latest SNe Ia Pantheon sample \cite{40} and BICEP2/KECK Array 2014 (BK14) combined polarization data \cite{41}. To explore the bound on $c$, on should completely consider the cases of early and late universe. For DE, we constrain LPSIDE using the combined datasets TTTEEE+lowP+lensing+BAO+Pantheon (hereafter TLLBP). For inflation, we constrain LPSIDE using the data combination of TTTEEE+lowP+lensing+BAO+BK14 (hereafter TLLBK).

\begin{table*}[!t]
	\renewcommand\arraystretch{1.5}
	\caption{For dark energy analysis, using the combined datasets TTTEEE$+$lowP$+$lensing$+$BAO$+$Pantheon (TLLBP), marginalized constraints on the model parameters of the $\Lambda$CDM, LPSCPL and LPSIDE models are shown, respectively. For inflation analysis, marginalized constraints on the model parameters of the LPSIDE model are also presented by using a data combination of TTTEEE$+$lowP$+$lensing$+$BAO$+$BK14 (TLLBK). The symbol `` $\diamondsuit$ '' denotes the parameter that cannot be well constrained by data in a given model. Note that we quote $1\sigma$ ($68\%$) errors for all the parameters except for the sum of three active neutrinos $\Sigma m_\nu$ and the tensor-to-scalar $r$ calculated at pivot scale $k=0.002$ Mpc$^{-1}$, for which we both quote $2\sigma$ ($95\%$) and $3\sigma$ ($99\%$) errors. Meanwhile, we quote $1\sigma$, $2\sigma$ and $3\sigma$ errors for the CPL parameter pair $(\omega_0,\,\omega_a)$.                  }
	\begin{tabular} { l  c c c c}
		\hline
		\hline
		
		Data               &   \multicolumn{1}{|c}{}    &   TLLBP    &        &  \multicolumn{1}{|c}{TLLBK}                                   \\
		\hline
		Model              &  \multicolumn{1}{|c|}{$\Lambda$CDM}                &\multicolumn{1}{c|}{LPSCPL}                        & LPSIDE       &\multicolumn{1}{|c}{LPSIDE}  \\
		\hline
		{\boldmath$\Omega_b h^2   $} & $0.02228\pm 0.00020 $     & $0.02238\pm 0.00064 $    &$0.02238\pm 0.00063        $    & $0.02227\pm 0.00033        $               \\
		
		{\boldmath$\Omega_c h^2   $} & $0.1182\pm 0.0011   $    & $0.128\pm 0.011    $    & $0.1240^{+0.0073}_{-0.0110} $  & $0.1208\pm 0.0037          $                                      \\
		
		{\boldmath$100\theta_{MC} $} & $1.04109\pm 0.00041  $   & $1.04024\pm 0.00082     $  & $1.04039\pm 0.00073        $  & $1.04090\pm 0.00051        $                                         \\
		
		{\boldmath$\tau           $} & $0.069\pm 0.013   $     &$0.086^{+0.020}_{-0.018}   $   & $0.074^{+0.025}_{-0.022}   $   & $0.073\pm 0.024            $                                   \\
		
		{\boldmath${\rm{ln}}(10^{10} A_s)$} & $3.067\pm 0.024 $  & $3.116\pm 0.042    $    & $3.085\pm 0.048       $   & $3.079\pm 0.051            $                                        \\
		
		{\boldmath$n_s            $} & $0.9687\pm 0.0043  $    & $0.985^{+0.051}_{-0.046}   $  & $0.971^{+0.041}_{-0.046}   $  & $0.969^{+0.023}_{-0.021}   $                                        \\
		
		{\boldmath$\Omega_K       $} & ---       & $0.0071^{+0.0081}_{-0.0071}$   & $0.0113\pm 0.0093  $   & $0.0086^{+0.0089}_{-0.0110} $                        \\
		
		{\boldmath$\Sigma m_\nu   $}  & ---      & $< 0.917   $ $(< 1.12)$ & $< 0.926   $ $(< 1.23)$     & $< 0.602   $  $(< 0.742 )$             \\
		
		{\boldmath$w_0            $}  & ---      & $-1.063^{+0.094+0.21+0.31}_{-0.110-0.20-0.25}  $  & $-1.099^{+0.098+0.22+0.34}_{-0.120-0.20-0.24}   $  & $-0.67^{+0.22+0.83+1.10}_{-0.43-0.61-0.71}     $                       \\
		
		{\boldmath$w_a            $}  & ---      & $0.32^{+0.60+0.85+0.97}_{-0.34-1.10-1.50}   $  & $0.50^{+0.60+0.75+0.82}_{-0.28-1.10-1.50}  $  & $-0.74^{+1.70+2.0+2.2}_{-0.62-3.3-4.4}      $                       \\
		
		{\boldmath$N_{eff}        $}  & ---      & $3.8\pm 1.0$   & $3.45^{+0.69}_{-1.00}  $    & $3.18^{+0.31}_{-0.27}      $                \\
		
		{\boldmath$Y_{He}         $}   & ---     & $\diamondsuit$ & $\diamondsuit$     & $\diamondsuit$                      \\
		
		{\boldmath$\alpha    $}         & ---     &  $0.0002\pm 0.0041   $   & $-0.0019^{+0.0054}_{-0.0038}$   & $0.00079^{+0.00055}_{-0.00090}$                                  \\
		
		{\boldmath$A_{L}          $}    & ---   & $1.135^{+0.084}_{-0.130}  $  & $1.22^{+0.12}_{-0.20}      $   & $1.122^{+0.089}_{-0.170}    $                                      \\
		
		{\boldmath$n_{\rm run}    $}    & ---   & $0.029\pm 0.028   $   & $0.015\pm 0.024            $      & $0.009\pm 0.012            $                            \\
		
		{\boldmath$n_{\rm run, run}$}   & ---   & $0.040\pm 0.026  $  & $0.029^{+0.025}_{-0.031}   $  & $0.027\pm 0.016            $                         \\
		
		{\boldmath$r              $}    & ---   & $< 0.232 \, (0.281)  $    & $< 0.328 \, (0.438)  $    & $< 0.138 \, (0.167)  $                                          \\
		
		{\boldmath$\epsilon       $}    & ---    & ---   & $-0.00013\pm 0.00038 $    & $-0.00025\pm 0.00031       $                                \\
		
		\hline
		
		{\boldmath$H_0                       $ }& $67.99\pm 0.52             $ & $71.4\pm 3.8               $   & $70.0^{+2.9}_{-3.8}        $    & $65.3^{+3.2}_{-2.7}        $                         \\
		
		{\boldmath$\Omega_m                  $ }& $0.3053\pm 0.0067          $  & $0.306\pm 0.014            $    & $0.309\pm 0.014            $   & $0.344^{+0.023}_{-0.033}   $                                                       \\
		
		{\boldmath$\sigma_8                  $ }& $0.8159\pm 0.0091          $  & $0.765\pm 0.053            $      & $0.732^{+0.061}_{-0.071}   $   &  $0.738\pm 0.050            $                                                     \\
		\hline
		\hline
	\end{tabular}
	\label{t2}
\end{table*}

\section{Results}
Our constraining results on the CPL parameter pair ($\omega_0$, $\omega_a$) using TLLBP are shown in Fig.\ref{f1}.
To find the upper bound of $\omega(z)$ in LPSIDE, we calculate the maximal values of $\omega(z)$ along the $2\sigma$ and $3\sigma$ contours at each point lying in the range $z\in[0,1]$. To obtain the upper bound on $c$, we shall first estimate the order of magnitude of $Q$ and then solve Eqs.(\ref{3}-\ref{6}) numerically via the initial conditions $x_1\approx x_2\approx x_3\approx0$ and $x_4=0.999$ \cite{42}. Since $Q\rho_{dm}\dot{\phi}=[(QH\rho_{dm})/(1+z)](\mathrm{d}\phi/\mathrm{d}a)\approx\delta H\rho_{dm}$, combining our latest constraint on IDE parameter $\epsilon=-0.00013\pm0.00038$ in the multi-feature universe with well approximate conclusion $\mathrm{d}\phi/\mathrm{d}a\approx0.5$ derived from the upper left panel of Fig.3 in \cite{43}, we can easily find $Q\sim \mathcal{O}(10^{-3})$. Consequently, we fix $Q=0.001$ in this study. After some calculations, in Fig.\ref{f2}, we find that current constraints allow IQDE models with $c\lesssim1.62$ and $c\lesssim1.94$ at the $2\sigma$ and $3\sigma$ levels, respectively. Furthermore, in two fiducial models using TLLBP, we exhibit the $2\sigma$ ($3\sigma$) upper bounds on $c$ in Tab.\ref{t1} when $Q=0.001$ in the IQDE comparison model. We find that: (i) The $\Lambda$CDM is still consistent with our two multi-feature cosmological models, LPSCPL and LPSIDE, at the $1\sigma$ level; (ii) There is no signature of interaction between DM and DE in the dark sector of the multi-feature universe; (iii) The $2\sigma$ bound $c\lesssim1.62$ in LPSIDE increases surprisingly by a factor of 2.7 relative to $c\lesssim0.6$ obtained in \cite{29}; (iv) The $2\sigma$ and $3\sigma$ bounds on $c$ in LPSIDE have just $1\%$ increasement relative to those in LPSCPL. However, we should not belittle this small $1\%$ increasement, because it represents the effect of existence of interaction in the dark sector in the multi-feature fiducial cosmology on the constructions of possible string theory models of DE.

More interestingly and importantly, our $3\sigma$ bound $c\lesssim1.94$ allows the existences of many string theory models of DE ignoring vacuum stabilization. For example, the 11-dimensional supergravity based on the hyperbolic compactification with a double-exponential potential predicts the first exponent $\lambda_1\approx1.6$ \cite{25}, and the $O(16)\times O(16)$ heterotic string \cite{44,45}, a non-supersymmetric model constructed by twisting the $E_8\times E_8$ theory, also gives the smallest effective value $c\approx1.6$ \cite{25}. Moreover, we point out that all the Type IIA/B string models summarized in Table 1 in \cite{25} satisfy the bound $c\lesssim1.94$ would be favored by our constraint from TLLBP in the LPSIDE-IQDE comparison setting. Additionally, it is noted that in \cite{25} the null energy condition can also predict reasonable bounds on $c$ for different string compactifications.

For inflation analysis, we obtain the $2\sigma$ ($3\sigma$) upper bound on the tensor-to-scalar ratio $r<0.138$ $(0.167)$ by using the combined datasets TLLBK to constrain LPSIDE. The corresponding constraining results are shown in Fig.\ref{f3}. Being different from Planck-2018 prediction in $\Lambda$CDM \cite{22}, we find that our constraint allow the existences of hybrid model driven by logarithmic quantum corrections in spontaneously broken supersymmetric (SB SUSY) theories \cite{48}, power-law inflation, and $\phi^2$ inflation \cite{8} at the $1\sigma$, $2\sigma$ and $3\sigma$ levels, respectively. Naturally, the $R^2$ inflation \cite{2} is still well supported by current data. According to our constraint $n_s=0.969^{+0.023}_{-0.021}$, we find that the scale invariance of primordial power spectrum ($n_s=1$) can be well satisfied at less than $2\sigma$ level in LPSIDE.

Using the relation $r>8c^2$ from previous analysis \cite{49,50}, for the case of concave potential, we obtain the $2\sigma$ ($3\sigma$) bound $c<0.13$ $(0.14)$, which is basically consistent with $c<0.09$ obtained in \cite{29}. Clearly, one can easily find that our constraint also permits the existences of a number of models with convex potentials. Recently, Kinney {\it et al}. \cite{31} notice that non-canonical extensions of models with convex potentials can be brought into agreement with data.
Using Planck primordial non-Gaussianity and inflation constraints \cite{51}, they obtain a $2\sigma$ bound $c\lesssim0.37$ in the Dirac-Born-Infeld (DBI) inflationary model with convex potential, which is still in strong tension with \emph{SC2} \cite{31}. Interestingly, if using our $2\sigma$ ($3\sigma$) limit $r<0.138$ $(0.167)$ in LPSIDE, we can obtain the $2\sigma$ ($3\sigma$) constraint $c\lesssim0.53$ $(0.58)$, which is now in a modest tension with the swampland conjecture.

For dark energy analysis, we constrain the the $\Lambda$CDM, LPSCPL and LPSIDE models using the combined datasets TLLBP. To implement the inflation analysis, we constrain the LPSIDE model using a data combination TLLBK. The corresponding results are presented in Tab.\ref{t2} and Figs.\ref{f4}-\ref{f5}. Besides the above results, we also obtain three interesting conclusions: (i) The scale invariance of primordial power spectrum, which predicts the scalar spectral index $n_s=1$, can be well satisfied at less than $2\sigma$ confidence level in the LPSIDE model \cite{22}; (ii) The constraints on consistency parameter $A_L$ scaling the lensing spectrum in the multi-feature LPSIDE and LPSCPL models is now well consistent with the theoretical prediction $A_L=1$ at about the $1\sigma$ confidence level \cite{22}; (iii) In light of current data, we cannot well constrain the primordial helium abundance $Y_{He}$ (see Tab.\ref{t2}). It is very difficult to understand this `` helium puzzle '', since accurate measurement of $Y_{He}$ is one of the main predictions of hot big bang cosmology and has also been well verified by the Planck satellite \cite{22}. To a large extent, we think that this puzzle may be ascribed to unknown correlations between different cosmological parameters. We plan to explore this `` helium puzzle '' found in the multi-feature cosmology in the forthcoming study. Besides the above results, we do not find any departure from the analysis of Planck-2018 final release \cite{22}. We expect that, in the future, more high-quality data can place more precise constraints on the multi-feature cosmological scenarios.

\section{Discussions and Conclusions}
Based on the constructions of string theories, the swampland conjecture has recently been proposed \cite{24,25}. We argue that this conjecture is not just a simple generalization of no de Sitter property in the landscape and it may have more profound implications than what we could expect. In \cite{32}, the authors confront a 5-parameter quintessence dark energy model with data and find the $3\sigma$ bound $c\lesssim1.01$. Although their procedure does not depend on the swampland criteria, their result conversely verifies the validity of this conjecture.

In this study, assuming the validity of the swampland conjecture, we explore the possibly largest upper bound on the swampland constant $c$. Under the belief that the universe should be multi-feature and informative, choosing interacting quintessence dark energy as comparison model, we obtain the $3\sigma$ bound $c\lesssim1.94$ in the large parameter space interacting dark energy model, a 12-parameter extension to the $\Lambda$CDM model. This bound would permit the existences of a number of string theory models of dark energy such as 11-dimensional M-theory with double-exponential potential, $O(16)\times O(16)$ heterotic string and some Type II string compactifications. For inflationary models with concave potential, we find the $3\sigma$ bound $c\lesssim0.14$ similar to previous result $c\lesssim0.09$ \cite{29}. Interestingly, combining Planck primordial non-Gaussianity with inflation constraints, we find that the DBI inflation with concave potential gives the $3\sigma$ bound $c\lesssim0.58$, which is now in a mild tension with the swampland conjecture. Furthermore, the scale invariance of primordial power spectrum predicting the scalar spectral index $n_s=1$ can be well satisfied at less than $2\sigma$ confidence level in the large parameter space cosmology. Meanwhile, the value of consistency parameter $A_L$ scaling the lensing spectrum in the multi-feature cosmological models is now well consistent with the theoretical prediction $A_L=1$ at about the $1\sigma$ confidence level.

\section{Acknowledgements}
Deng Wang thanks Pengjie Zhang and Bin Wang for helpful discussions.

\end{document}